\documentclass[%
reprint,
superscriptaddress,
 amsmath,amssymb,nofootinbib,
 aps,
 twocols
]{revtex4-1}

\usepackage[title]{appendix}
\usepackage{verbatim}
\usepackage{graphicx}% Include figure files
\usepackage{dcolumn}% Align table columns on decimal point
\usepackage{bm}
\usepackage{mathtools}
\usepackage{float}
\makeatletter
\let\newfloat\newfloat@ltx
\makeatother
\usepackage{algorithm} 
\usepackage{algpseudocode}  
\usepackage{float}                                    %%%
\usepackage{graphicx} 
\usepackage[font=small,labelfont=bf,justification=raggedright,format=plain]{caption}
\usepackage{hyperref}                                 %%%
\usepackage[usenames, dvipsnames]{color}
\usepackage{color}
\usepackage{listings}
\usepackage{enumerate}
\newcommand{\map}[0]{{\bf M}}

\definecolor{dkgreen}{rgb}{0,0.6,0}
\definecolor{gray}{rgb}{0.5,0.5,0.5}
\definecolor{mauve}{rgb}{0.58,0,0.82}

\begin{document}

%\preprint{APS/123-QED}

\title{Coarse-grained Mori-Zwanzig dynamics in a time-non-local stationary-action framework}

%breaks with \\

%\author{Aaa Bbb}
\author{Piero Luchi}
\thanks{These authors contributed equally to this work.}
\affiliation{Physics Department, University of Trento, via Sommarive, 14 I-38123 Trento, Italy}
\affiliation{INFN-TIFPA, Trento Institute for Fundamental Physics and Applications, I-38123 Trento, Italy}
\author{Roberto Menichetti}
\thanks{These authors contributed equally to this work.}
\affiliation{Physics Department, University of Trento, via Sommarive, 14 I-38123 Trento, Italy}
\affiliation{INFN-TIFPA, Trento Institute for Fundamental Physics and Applications, I-38123 Trento, Italy}
\author{Gianluca Lattanzi}%
\affiliation{Physics Department, University of Trento, via Sommarive, 14 I-38123 Trento, Italy}
\affiliation{INFN-TIFPA, Trento Institute for Fundamental Physics and Applications, I-38123 Trento, Italy}\author{Raffaello Potestio}%
 \email{raffaello.potestio@unitn.it}
\affiliation{Physics Department, University of Trento, via Sommarive, 14 I-38123 Trento, Italy}
\affiliation{INFN-TIFPA, Trento Institute for Fundamental Physics and Applications, I-38123 Trento, Italy}


%\affiliation{
%\affiliation{
%\affiliation{%

\date{\today}% It is always \today, today,

\begin{abstract}
Coarse-grained (CG) models are simplified representations of soft matter systems that are commonly employed to overcome size and time limitations in computational studies. Many approaches have been developed to construct and parametrise such effective models for a variety of systems of natural as well as artificial origin. However, while extremely accurate in reproducing the stationary and equilibrium observables obtained with more detailed representations, CG models generally fail to preserve the original time scales of the reference system, and hence its dynamical properties. In order to improve our understanding of the impact of coarse-graining on the model system dynamics, we here formulate the Mori-Zwanzig generalised Langevin equations (GLEs) of motion of a CG model in terms of a time non-local stationary-action principle. The latter is employed in combination with a data-driven optimisation strategy to determine the parameters of the GLE. We apply this approach to a system of water molecules in standard thermodynamical conditions, showing that it can substantially improve the dynamical features of the corresponding CG model.
\end{abstract}

\maketitle

%\tableofcontents

In the computational investigation of soft and biological matter, a steadily increasing use is made of coarse-grained (CG) models, that is, simplified representations of a system in which a group of atoms is lumped into a pseudo-atom, or {\it bead}, which interacts with the others through effective potentials \cite{rudzinski2011coarse,saunders2013coarse,potestio2014computer,d2015coarse,giulini2021system}. In comparison to accurate but resource-intensive all-atom descriptions, these models enable the long-time simulation of large systems, and offer the advantage of a relatively small number of degrees of freedom to analyse. A broad spectrum of methods has been developed to construct CG models; in bottom-up strategies, in particular, the effective interactions are parametrised through exact or approximate integration of the reference system's degrees of freedom \cite{noid2008multiscale,rudzinski2011coarse,shell2016coarse,lebold2019dual,dannenhoffer2019compatible,giulini2021system}.

\begin{figure}[h]
\includegraphics[scale=0.25]{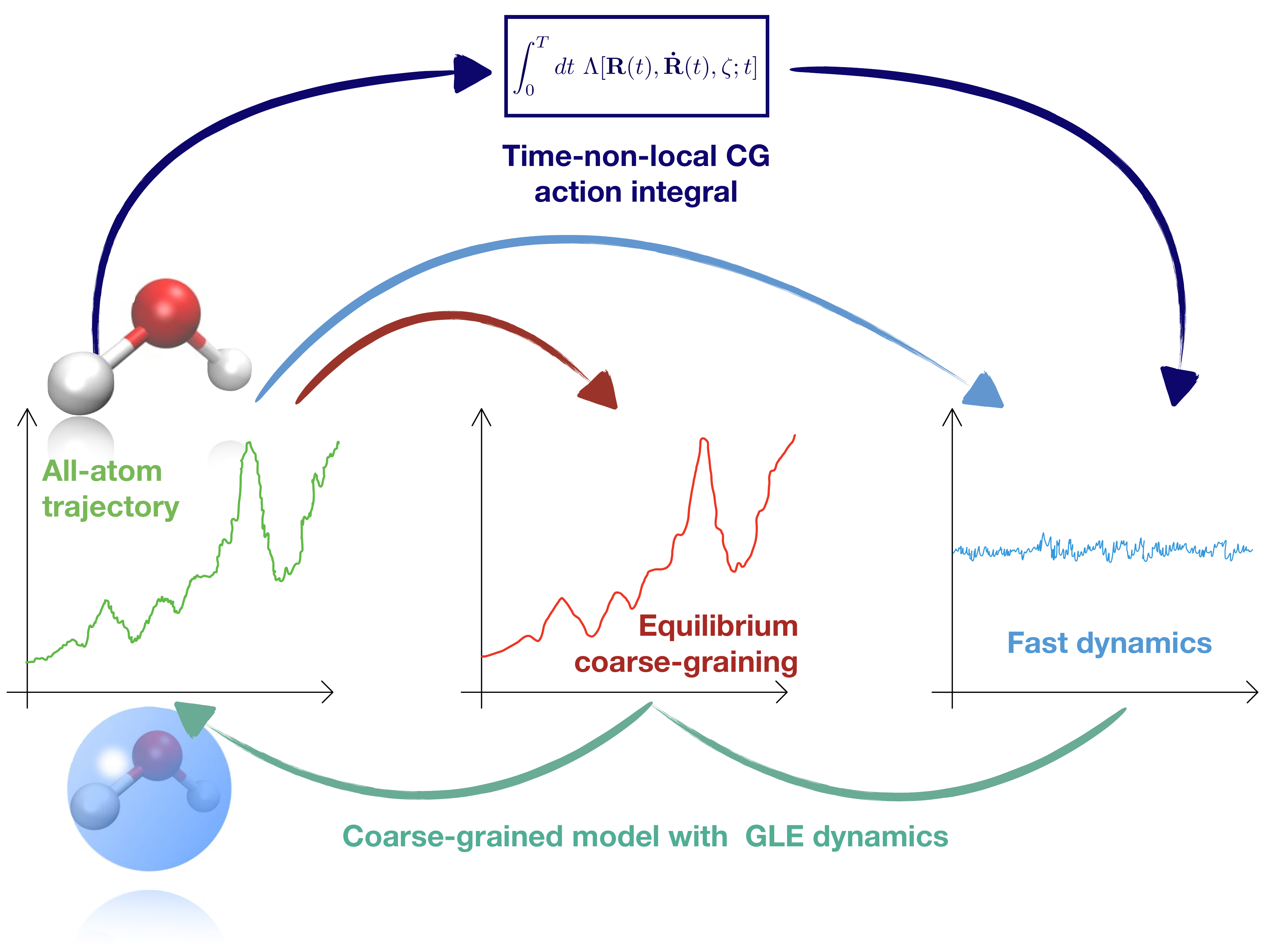}
\caption{Schematic illustration of the method developed and employed in this work. The atomistic trajectory of the all-atom water system, following Newton's equation of motion (first panel), is mapped onto its coarse-grained trajectory (second panel). The fluctuations of the fast dynamics, which are integrated out in this procedure, are depicted in the third panel. The method presented in this work aims to minimise a coarse-grained time-non-local action with respect to the fast dynamics parameters in order to recover them as noise and memory terms of a generalised Langevin equation. Finally, the numerical integration of the latter shows a recovery of the dynamics of the original all-atom system.}
\label{fig:graphical_abstract}
\end{figure}

A substantial limitation that is intrinsic in the process of coarse-graining affects its capability to reproduce the system's {\it characteristic time scales} \cite{rudzinski2019recent,klippenstein2021introducing,izvekov2021mori,schilling2021coarse}. In fact, a correct conformational sampling at equilibrium does not imply an equivalently accurate reproduction of bond vibrations frequencies, molecular diffusion, or the time required by a local deformation to propagate throughout the structure. In general, the dynamics of a coarse-grained system is accelerated and distorted with respect to that of the reference, higher resolution model, due to the fewer degrees of freedom, the softer potentials, and the consequently smoother (free-)energy landscape; for classes of problems ranging from the diffusion of particles in a solvent \cite{van2007coarse,shin2010brownian,davtyan2015dynamic,lei2016data,jung2017iterative,jung2018generalized,bockius2021model} up to the analysis of the behavior of complex molecular fluids \cite{izvekov2006modeling,hijon2010mori,davtyan2015dynamic,li2015incorporation,li2017computing,yoshimoto2017construction,han2018mesoscopic,wang2019implicit,wang2020data}, several works have investigated the connection between the underlying reference system and the emergent CG dynamics \cite{rudzinski2019recent,schilling2021coarse,klippenstein2021introducing,izvekov2021mori}.

One of the most rigorous---and pioneering---bottom-up frameworks describing the relationship between a high-resolution model, its coarse-grained interactions, and the corresponding low-resolution model dynamics is provided by the Mori-Zwanzig (MZ) formalism \cite{Zwanzig_book,darve2009GenLagevi,di2019systematic,schilling2021coarse}. In this picture, the Newton's equations of motion of the atomistic system are projected onto a set of generalised Langevin equations (GLE's) for the CG degrees of freedom; in addition to the effective conservative forces acting among the CG sites, memory and noise terms appear in the GLE that implictly account for the effect generated on the CG dynamics by the orthogonal, fast degrees of freedom. Critically, the complexity of extracting such non-conservative components has for long time limited the applicability of the MZ formalism, and only recently approaches have been introduced that aim at directly tackling their parametrisation to construct dynamically consistent CG models. Specifically, in few cases the memory and noise terms are obtained via analytical techniques \cite{chen2014computation,ma2016derivation}. More often, an all-atom calculation via computer simulations of a set of CG correlation functions is performed for the system of interest; subsequently, the relations linking these ingredients with the fundamental GLE ones---e.g., the Volterra equations linking the force-velocity and velocity-velocity correlation functions to the memory kernel---are inverted via direct \cite{izvekov2006modeling,li2015incorporation,lei2016data,li2017computing,yoshimoto2017construction,han2018mesoscopic,wang2019implicit,bockius2021model}, iterative \cite{jung2017iterative,jung2018generalized} or machine learning techniques \cite{wang2020data} so as to determine the non-conservative factors of the GLE.

\begin{figure}
\includegraphics[scale=0.06]{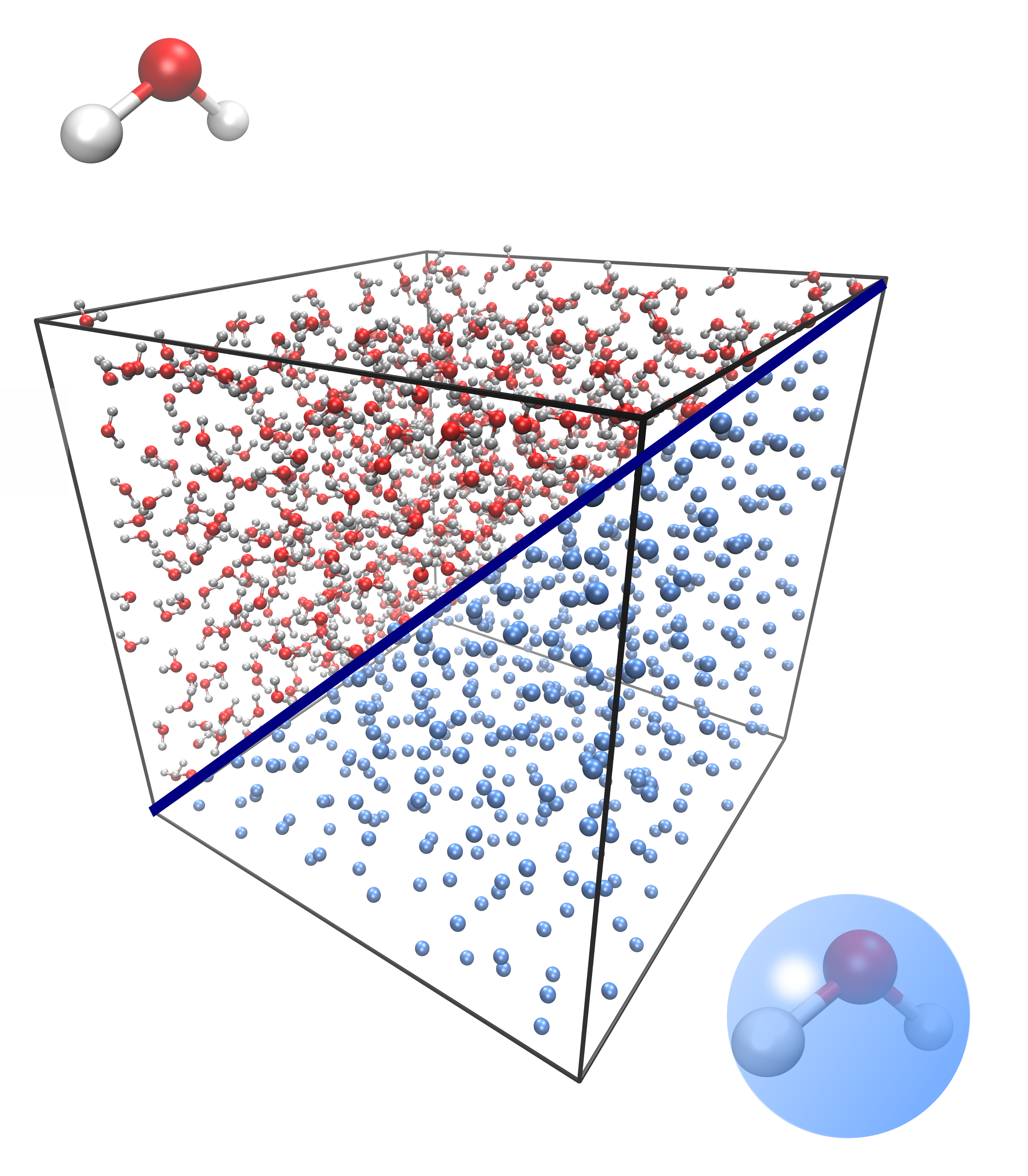}
\caption{Representation of the water system employed in this work. On the left part the system is represented with atomistic resolution, while on the right side a coarse-grained model is used, in which a water molecule is treated as a single CG bead.}
\label{fig:water_system}
\end{figure}

As anticipated, the aforementioned approaches build upon the Mori-Zwanzig formalism; this, in turn, assumes a Hamiltonian framework for the dynamics of the underlying high-resolution system, while, as a consequence of a projection procedure, the dynamics of the CG model emerges as a set of non-conservative GLE equations. In this work we explore the possibility of obtaining the dynamics of both the reference, high-resolution system and that of the low-resolution CG one from a stationary-action principle; taking advantage of this common framework, we establish a bridge that connects the time evolution of the system in these two representations and, at the same time, provides a practical, viable strategy to parametrise the noise and memory terms of the CG model's GLEs.
The method is applied to a system of water molecules: Fig.~\ref{fig:graphical_abstract} schematically illustrates the procedure, while Fig.~\ref{fig:water_system} highlights the CG representation of water employed to in this work, with each molecule being mapped onto a CG bead located on the molecule's center of mass.

\section{Structural and dynamical consistency between models at different resolution}\label{sect:hypoth_theor_back}

The simulation of an arbitrary atomistic system relies on the numerical integration of its Newton's equations of motion,
\begin{eqnarray} \label{NewtonEOM}
m_i \ddot{\mathbf{r}}_i=\mathbf{F}_i \ \ \ \mathrm{with}\ \ \ \mathbf{F}_i=-\frac{\partial u(\bf r) }{\partial \mathbf{r}_i},
\label{eq:newton_eq}
\end{eqnarray}
where ${\bf r}_i$, $i=1,...,n_A$ are the Cartesian coordinates of the constituent atoms in the system and  $u({\bf r})$ is the potential energy taking into account non-bonded (e.g. van der Waals, electrostatic...) as well as bonded (e.g. bonds, angles, dihedrals...) interactions \cite{gonzalez2011AAforcefield_rev,fuentes2015AAforcefield}.

The MZ formalism rephrases these equations in terms of a set of collective, CG variables ${\bf R}_I$, $I=1,...,N_B<n_A$ undergoing a {\it slow} dynamics. The ${\bf R}_I$ are related to the original coordinates through a vector-valued {\it mapping function} $\map({\bf r})$, namely
\begin{eqnarray}\label{eq:mapping}
&&\mathbf{R}_I = \map_I(\mathbf{r}) = \sum_{k\in S_I}\frac{m_k}{M_I} \mathbf{r}_k, \\ \nonumber
&&\mathbf{P}_I = \sum_{k\in S_I}\mathbf{p}_k,
\end{eqnarray}
where ${\bf R}_{I},\ {\bf P}_I$ and $M_I = \sum_{k \in S_I} m_k$ are respectively the coordinates, momenta, and masses of the $I^{th}$ CG bead, ${\bf r}_k$, ${\bf p}_k$  and $m_k$ are respectively the coordinates, momenta, and masses of the $k^{th}$ atom, and $S_I$ is the set of atoms that map onto bead $I$.

This mapping, together with further manipulations  and simplifications of Eq.~\ref{NewtonEOM} \cite{schilling2021coarse,di2019systematic,hijon2010mori} (see Supporting Information for a short recap), yields the generalised Langevin equations for the time evolution of the CG degrees of freedom, which read:
\begin{eqnarray}\label{eq:MZ_simpl_GLE}
&&\dot{\bf R}_I = \frac{\bf P_I}{M_I}, \\ \nonumber
&&\dot{\bf P}_I = -\frac{\partial U({\bf R})}{\partial  \bf R_I}-\int_{0}^{t}dt' K(t-t^\prime) \frac{\bf P_I}{M_I}  + {\bm \xi}_I(t),
\end{eqnarray} 
where $U({\bf R})$ is the effective potential of the CG system induced by the mapping, while $K(\tau) $ is defined as the memory kernel.  The term $\bm \xi(t)$ originates from the projection of the fast degrees of freedom onto a rapidly-varying force acting on the slow, CG coordinates, and is typically modelled as an autocorrelated noise linked to the memory kernel $K(\tau)$ {\it via} the fluctuation-dissipation theorem~\cite{kubo1966FDT},
\begin{eqnarray}\label{FDT}
&&\langle  \xi^{\alpha}_I(t+\tau) {\xi}^{\beta}_J(t) \rangle = k_BT K(\tau) \delta^{\alpha\beta}\delta_{IJ},
\end{eqnarray}
where the indices $I,J=1,...,N_B$ run over the CG particles while $\alpha,\beta=x,y,z$ over the spatial dimensions.

The first step in the parametrisation of a CG model according to Eq.~\ref{eq:MZ_simpl_GLE} is the determination of the effective CG potential $U(\bf R)$. This task is pursued by systematic coarse-graining \cite{noid2008multiscale,potestioEntropy2014,giulini2021system}, which aims at developing potentials acting among effective interaction sites, typically representative of groups of atoms as in Eq.~\ref{eq:mapping}. The objective of this approach is that the equilibrium probability distributions of structural observables (e.g. two-body correlation functions), computed on the ensemble of CG structures sampled by the model, coincide with those obtained from the higher-resolution, all-atom structures {\it observed in terms of the CG degrees of freedom} \cite{noid2008multiscale,rudzinski2011coarse,shell2016coarse,lebold2019dual,dannenhoffer2019compatible,giulini2021system}. Irrespectively of the method, algorithm, or instrument employed to parametrise the interactions, the aforementioned objective provides not only a guideline in the construction of coarse-grained models, but also a quantitative criterion to perform such parametrisation and to assess the validity of its result.

In fact, quantitative agreement between structural correlations (in principle at arbitrary order) among the CG degrees of freedom of the reference and those of the model can be achieved if and only if a set of {\it consistency conditions}~\cite{noid2008multiscale} is satisfied, these being:

\begin{enumerate}
\item Each CG coordinate $\{\mathbf{R,P}\}$ is a well defined linear combination of a subset of the underlying atomistic coordinates $\{\mathbf{r,p}\}$, as it is the case in Eq.~\ref{eq:mapping};
\item The CG equilibrium distribution $P_{R}(\mathbf{R})$ associated to the effective potential $U(\bf R)$ is equal to the atomistic equilibrium distribution after mapping, $p_{R}(\mathbf{R})$. 
\end{enumerate}
Assuming a system at thermal equilibrium in the canonical ensemble, $p_{R}(\mathbf{R})$ is defined as
\begin{equation}
p_{R}(\mathbf{R})=\frac{1}{Z}\int d{\bf r} \ e^{-\beta u(\bf{r})}  \delta( \map({\bf r})-{\bf R}),
\end{equation}
where $Z$ is the canonical configurational partition function. A CG potential $U({\bf R})$ correctly reproduces the reference system's structural properties if it is equal to the {\it multi-body potential of mean force} (MB-PMF) $W({\bf R})$, that is 
\begin{eqnarray}\label{eq:consist:a}
U({\bf R})=W({\bf R}) = -k_BT\ln[p_{R}(\mathbf{R})]+\emph{const}.
\end{eqnarray}
Although exactly determining the MB-PMF is an extremely difficult task, Eq.~\ref{eq:consist:a} can be considered the constitutive equation of coarse-graining, in that it provides an exact prescription of what a CG model is required to satisfy in order to entail all those features of the underlying high-resolution model that can be preserved at that level of resolution. This, however, only pertains {\it structural} properties at equilibrium. Recently, Di Pasquale and coworkers have shown~\cite{di2019systematic} that the parametrisation of the CG interaction potentials and that of the GLE-related quantities (the memory kernel and the noise term in Eq.~\ref{eq:MZ_simpl_GLE}) are independent from each other: this is a key property that allows one to solve the structural problem by fixing the interaction potentials with some standard coarse-graining procedure, and subsequently fine-tune the memory and noise terms that are responsible for the dynamical behaviour of the system \cite{izvekov2006modeling,li2015incorporation,li2017computing,yoshimoto2017construction,han2018mesoscopic,wang2019implicit}.

To parametrise these terms we here turn our attention to the Euler-Lagrange formalism, or more appropriately to the stationary action principle, which allows us to characterise the global properties of a \emph{single trajectory as a whole}. We assume the underlying reference, high-resolution, or for short {\it atomistic} system to be composed by point-like particles that follow classical (Newton's) equations of motion in the {\it microcanonical} ensemble; as such, the system can be described {\it via} an action functional that is minimal in correspondence of the trajectory effectively followed by the system. Making use of the Mori-Zwanzig formalism, this trajectory is projected onto a set of slow, collective variables subject to a reduced (coarse-grained) force field plus a friction and a noise term; the latter are in general not {\it Langevin-like}, in that the noise is not delta-correlated, but rather it has a slowly-decaying, finite-time correlation function that is proportional to the memory kernel of the friction term. Conversely, the properties of the dynamics as observed in terms of the slow collective variables have to be consistent with the underlying microcanonical time evolution. We thus assume that the same structure in terms of a stationary-action principle can be preserved to describe the time evolution of the slow variables, provided that the interactions and the friction/noise terms are appropriately parametrised. Relying on these assumptions, we hereafter describe a consistency condition for the dynamics, and test its validity and effectiveness.

\section{Euler-Lagrange formulation of dynamical consistency}\label{sect:formulation}

In the Lagrangian formulation of classical mechanics, the equations of motion of the system can be derived from the minimisation (or extremisation) of the action functional $S$:
\begin{eqnarray}\label{eq:euler-lagr}
&&S[{\bf r}(t), \dot{\bf r}(t)] = \int_0^T dt \ L(t) = \int_0^T dt\ L({\bf r}(t), \dot{\bf r}(t)),\\ 
&&L({\bf r}(t), \dot{\bf r}(t)) = K(\dot{\bf r}(t)) - u({\bf r}(t)),\\ 
\label{eq:euler_lagr_newt}
&&\delta S = 0 \Rightarrow \frac{d}{dt}\left(\frac{\partial L}{\partial \dot{\bf r}_i}\right) = \frac{\partial L}{\partial {\bf r}_i},
\end{eqnarray}
where $K = \sum^{n_A}_{i=1} m_i \dot{\bf r}_i^2 / 2$ is the kinetic energy of the constituent particles, $u({\bf r})$ is the interaction potential acting among them, and appropriate boundary conditions are imposed on the configuration of the system at time $t = 0$ and $T$.

Let us now assume that the coordinates $\bf r$ of the system can be separated, {\it via} a linear transformation akin to Eq. \ref{eq:mapping}, in two groups: the fast coordinates $\bf q$ and the slow coordinates $\bf Q$. The Lagrangian becomes
\begin{eqnarray}\label{eq:euler-lagr-sep-1}
&&L( {\bf Q}(t), \dot{\bf Q}(t), {\bf q}(t), \dot{\bf q}(t)) =  K_Q(\dot{\bf Q}(t))  + \\ \nonumber
&&+ K_q(\dot{\bf q}(t)) - U_Q({\bf Q}(t)) - u_q({\bf q}(t)) - u_c({\bf q}(t), {\bf Q}(t)),
\end{eqnarray}
where the first two terms are respectively the kinetic energies of the fast and slow variables, the second two terms pertain to the interactions {\it within} each of these two groups of coordinates, while the last term contributes the interaction energy between fast and slow degrees of freedom.

Based on this Lagrangian, the action functional $S$ in Eq.~\ref{eq:euler-lagr} can be written as (for notational simplicity we omit the time dependence of the fast and slow variables)
\begin{equation}\label{eq:euler-lagr-sep-2}
S[ {\bf Q}, \dot{\bf Q}, {\bf q}, \dot{\bf q}] =  S_Q[ {\bf Q}, \dot{\bf Q}] + S_q[{\bf q}, \dot{\bf q}] + S_c[{\bf q}, {\bf Q}],
\end{equation}
with
\begin{eqnarray}\label{eq:euler-lagr-sep-3}
&& S_Q[ {\bf Q}, \dot{\bf Q}] = \int_0^T dt\ \left[K_Q(\dot{\bf Q}) - U_Q({\bf Q}) \right],\\ 
&&S_q[{\bf q}, \dot{\bf q}] = \int_0^T dt\ \left[K_q(\dot{\bf q}) - u_q({\bf q})\right], \label{eq:euler-lagr-sep-4}\\ 
&&S_c[{\bf q}, {\bf Q}] = - \int_0^T dt\ u_c({\bf q}, {\bf Q}) \label{eq:euler-lagr-sep-5}.
\end{eqnarray}
Eqs.~\ref{eq:euler-lagr-sep-2}-\ref{eq:euler-lagr-sep-5} display that the minimisation of $S$ could be performed separately for the two sets of variables if the coupling term $S_c$ were zero: it is the presence of such term that makes the time evolution of the fast and slow degrees of freedom mutually dependent. On the other hand, we can now assume that slow variables follow the \emph{actual} trajectory that minimises the action, i.e. ${\bf Q}(t) = {\bf Q}^\star(t)$. By doing this, $S$ can be considered a functional of the trajectory ${\bf q}(t)$ of the fast variables only, namely
\begin{equation}\label{eq:euler-lagr-sep-6}
S[ {\bf Q}^\star, \dot{\bf Q}^\star,{\bf q},\dot{\bf q}] = S_Q[ {\bf Q}^\star, \dot{\bf Q}^\star] + S_q[{\bf q}, \dot{\bf q}]  + S_c[{\bf q}, {\bf Q}^\star]. \\
\end{equation}
The action $S$ is minimised when computed over the physical trajectory $({\bf q}^\star(t), {\bf Q}^\star(t))$ of both the fast and slow variables; however, since the potential and kinetic energies where ${\bf Q}$ appears are already calculated onto such trajectory, it is now sufficient to perform the minimisation of Eq.~\ref{eq:euler-lagr-sep-6} with respect to the path followed by the fast variables ${\bf q}$. In doing so, we note that the term $S_Q$ in Eq.~\ref{eq:euler-lagr-sep-6} is constant with respect to the fast variables, and can be thus ignored in the differentiation of the action with respect to them. The integrand of the coupling term $S_c$, on the other hand, is now a function of the sole ${\bf q}(t)$ and time, and represents a background field for the dynamics of the fast variables.

This rather general framework is now specialised and applied to a particular case, with the particular extension to a time non-local action functional. Specifically, we will identify the slow variables in the coarse-grained degrees of freedom, and assume that their time evolution ${\bf Q}(t)$ has been solved, so that the minimisation of the action with respect to the remaining fast variables will provide us with the complete picture.

Implementing the identification of the slow variables with the coarse-grained ones and making use of Eq.~\ref{eq:mapping}, we thus have
\begin{eqnarray} \label{GLE}
&&{\bf Q}_I(t)={\bf R}_I(t) = \map_I({\bf r}(t)),\\ \nonumber
&&\dot{\bf Q}_I(t)=\dot{\bf R}_I(t) = \map_I(\dot{\bf r}(t)),
\end{eqnarray}
where $I=1,...,N_B$ and $N_B$ is the number of CG particles. Subsequently, we introduce an action $\Sigma$ representing an \emph{effective} analogue of the action $S$ in Eq.~\ref{eq:euler-lagr-sep-2}:
\begin{eqnarray}\label{eq:Lagrangian_subs}
\Sigma[{\bf R},\dot{{\bf R}};{\bm \zeta}] &=& \int_0^T dt\ \left[K_R(\dot{\bf R}(t)) - U({\bf R}(t))\right] +\\ \nonumber
&+& \int_0^T dt\ A[{\bf R}, \dot{\bf R}; t,{\bm \zeta}].
\end{eqnarray}
Here, $K_{R}$ and $U$ are, respectively, the kinetic and (effective) potential energy of the CG particles, and the term $A[{\bf R}, \dot{\bf R}; t, {\bm \zeta}]$ couples these slow variables with the fast ones, the latter being now represented by \emph{a set of properties} ${\bm \zeta}=[\zeta_1,...,\zeta_M]$. The function $A$ thus recapitulates all the information contained in the two latter terms of the r.h.s. of Eq.~\ref{eq:euler-lagr-sep-2}.

We now assume the dynamics followed by the fast variables, and their impact on the slow ones, to fall among those cases that can be treated through the Mori-Zwanzig formalism. The functional form of the action $\Sigma$ can then be determined by requiring that the equations of motion one obtains through its minimisation with respect to the slow variables coincide with the GLE describing the dynamics of the CG system. Specifically, we assume that the time evolution $I$-th CG particle, all particles having the same mass $M$, is dictated by
\begin{eqnarray}\label{eq:CG-eom}
&&M \ddot{{\bf R}}_I = {\bf F}_I({\bf R}(t)) - \int_{0}^t dt' K(t - t^\prime, {\bm \zeta}) \dot{\bf R}_I(t^\prime) + {\bm \xi}_I(t,{\bm \zeta}),  \nonumber \\
&&{\bf F}_I({\bf R}(t)) = - \frac{\partial U({\bf R})}{\partial {\bf R}_I}\bigg\rvert_{{\bf R}(t)}, \
\end{eqnarray}
where the memory kernel $K(\tau,{\bm \zeta})$ and the noise ${\bm \xi}_I(t,{\bm \zeta})$ are now \emph{parametric functions} of the properties ${\bm \zeta}$ that account for the effect of the fast variables, and satisfy the fluctuation-dissipation relation in Eq.~\ref{FDT}.

The presence of the memory term in Eq. \ref{eq:CG-eom} imposes some care in the construction of the action $\Sigma[{\bf R},\dot{{\bf R}};{\bm \zeta}]$ in Eq.~\ref{eq:Lagrangian_subs}: indeed, as these depend on the history of all velocities of the CG particles up to the current time, $\Sigma$ has to be written as the integral of a time non-local function. Critically, in such a case the usual differentiation with respect to positions and velocities at fixed time that appears in the ``traditional'' Euler-Lagrange approach of Eq.~\ref{eq:euler_lagr_newt} can no longer be performed. To tackle this problem, Ferialdi and Bassi \cite{ferialdi2012TNL} recently developed a generalised framework in which the equations of motion generated by minimising a time non-local action $\Sigma$ are defined as 
\begin{eqnarray}\label{eq:TNL_Euler_Lagrange_eq}
\frac{\delta \Sigma [ {\bf R}, \dot{{\bf R}}; {\bm \zeta}]}{\delta {\bf R}_I(s)}-\frac{d }{d s}\frac{\delta \Sigma [  {\bf R}, \dot{{\bf R}}; {\bm \zeta}]}{\delta \dot{{\bf R}_I}(s)}=0,
\end{eqnarray}
where $s\in[0,T]$ and ${\delta}/{\delta {\bf R}(s)}$ and ${\delta}/{\delta \dot{{\bf R}}(s)}$ are now functional derivatives with respect to the positions and velocities of the $I$-th effective CG site, respectively. Note that, in the case of a time-local Lagrangian function, Eq. \ref{eq:TNL_Euler_Lagrange_eq} reduces to the usual form of Eq. \ref{eq:euler_lagr_newt}.

Based on this extension of the Euler-Lagrange formalism, we here propose an effective time non-local action $\Sigma$ associated to the GLE of the CG system:
\begin{eqnarray}\label{eq:CG_act_expl}
&\Sigma[{\bf R}, \dot{\bf R};{\bm \zeta}]  =  \int_0^T dt \left [  K_R(\dot{\bf R}(t)) - U({\bf R}(t))+  \right . \\
 &+ \left . \sum_{J=1}^{N_B} \dot{\bf R}_J(t)\cdot\int_{0}^t dt' \  Q(t - t',{\bm \zeta}) \dot{\bf R}_J(t') + \right . \nonumber \\
 & + \left .\sum_{J=1}^{N_B} \bm{\xi}_J(t,{\bm \zeta})\cdot{\bf R}_J(t)\right], \nonumber 
\end{eqnarray}
where $Q(\tau,{\bm \zeta})$ is related to the memory kernel $K$ in Eq.~\ref{eq:CG-eom} by $\frac{d}{d\tau}Q(\tau,{\bm \zeta})=K(\tau,{\bm \zeta})$. By plugging the action $\Sigma$ of Eq.~\ref{eq:CG_act_expl} in Eq.~\ref{eq:TNL_Euler_Lagrange_eq}, one obtains the generalised Langevin equations of motion presented in Eq.~\ref{eq:CG-eom} for the dynamics of the CG particles, under the requirement---representing an approximation---that the time evolution of the low-resolution system depends only on its past, and not on its future; for the sake of brevity we will here omit the full derivation, all details being provided in the Supporting Information. 

As anticipated, it is our assumption that the CG trajectory ${\bf R}(t)$ actually follows a dynamics for which the action $S$ in Eq.~\ref{eq:euler-lagr-sep-2} is extremal. Given the effective action $\Sigma[{\bf R}, \dot{\bf R}; {\bm \zeta}]$ defined in Eq.~\ref{eq:CG_act_expl}, we treat the trajectory of the CG variables as fixed, in that it is determined by the reference, all-atom coordinates ${\bf r}(t)$ projected onto the CG degrees of freedom through the mapping $\map({\bf r}(t))$, see Eq.~\ref{eq:mapping}. Subsequently, we look for the fast dynamics properties ${\bm \zeta}$ that minimise $\Sigma$, where the properties enter such action {\it via} its memory and noise components. This operation is justified by the assumption that the memory/noise terms of the GLE are actually representative of the fast degrees of freedom, so that minimising $\Sigma$ with respect to the properties ${\bm \zeta}$ is equivalent to minimising it with respect to the trajectories of those variables. Our \emph{consistency equation for the dynamics} of the CG model thus reads
\begin{eqnarray}\label{eq:const-dyn}
\frac{\partial \Sigma[{\bf R}, \dot{\bf R}; {\bm \zeta}]}{\partial {\bm \zeta}}\bigg\rvert_{\bm \zeta = \bm \zeta^{\star}} = 0.
\end{eqnarray}
The solution $\bm \zeta^{\star}$ to Eq.~\ref{eq:const-dyn} will provide \emph{optimised} parameters characterising the influence of the fast dynamics on the time evolution of the low-resolution system; in the case under consideration, these consist of optimised memory kernel and noise terms $K(\tau,{\bm \zeta^{\star}})$ and ${\bm \xi}_I(t,{\bm \zeta^{\star}})$. These ingredients can be subsequently employed for \emph{generating} the CG trajectory via the associated GLE, see Eq.~\ref{eq:CG-eom}, the statistical properties of which can be finally compared to those predicted by the original, mapped trajectory $\map({\bf r}(t))$ to quantify the accuracy of the approach.

In practice, the effective action $\Sigma$ in Eq.~\ref{eq:CG_act_expl} can be recast in the form of a sum over discrete times, enabling its calculation on the trajectory obtained by projecting the results of an atomistic molecular dynamics simulation of the system onto their CG counterpart. A parametric functional form of the memory kernel and the associated noise is imposed through an {\it Ansatz} based on physical considerations, see Sec.~\ref{sec:kern}; the minimisation in Eq. \ref{eq:const-dyn} is then carried out with respect to these parameters so as to determine their optimised values entering the GLE of the CG system. A set of constraints is further introduced in the minimisation protocol in order to stabilise the numerical solution to the particular optimisation problem that emerges from our Ansatz for the kernel; all details about these technical steps are provided in the Supporting information.

\section{Discussion}\label{sect:application}

The paramount importance of water for practically all natural and artificial processes, as well as its remarkable and unique physical properties \cite{eisenberg2005structure, tomobe2015_VAF}, has constantly fuelled the development of a vast library of atomistic and coarse-grained models of this molecule \cite{sanz2004phase,bizzarri2002molecular,anderson1995_CFD,olbers2012ocean,mark2001atomistic_water,johnson2007_water_model_simple,wu2010CG_water_model_complex,riniker2011_water_model_complex_2}. Given the structural simplicity and phenomenological complexity of water, we focussed on it to put our coarse-graining method at test. Specifically, we proceeded as follows: first, microcanonical all-atom simulations of a water system at an average temperature of $T=298$~K were performed, which provided the reference time evolution for the CG modelling. Subsequently, by mapping each water molecule onto a CG bead located in the molecule's center of mass, see Fig.~\ref{fig:water_system}, we moved to the construction of a force field for the low-resolution system. We relied on a pairwise-additive approximation, and determined the CG pair potential $U_2(R)$ between two water beads via the iterative Boltzmann inversion method \cite{reith2003IBI,rosenberger2016IBI} starting from the atomistic radial distribution function (RDF) $g(R)$ among the water molecules' centers of mass. Finally, we optimised the parameters of the GLE, namely the kernel $K(t,\bm \zeta)$ and the noise $\bm \xi(t,\bm \zeta)$ that appear in Eq.~\ref{eq:CG-eom}. As outlined in the previous section, this was achieved through the minimisation of the action $\Sigma[{\bf R}, \dot{\bf R};{\bm \zeta}]$ in Eq.~\ref{eq:CG_act_expl} w.r.t. the properties $\bm \zeta$, where $\Sigma$ was calculated over the all-atom trajectory of the water system projected onto the CG degrees of freedom.
An exhaustive description of the technical details involved in each of these steps is presented in Sec.~\ref{sec:methods} and in the Supporting Information.

\begin{figure*}[ht]
\includegraphics[width=\textwidth]{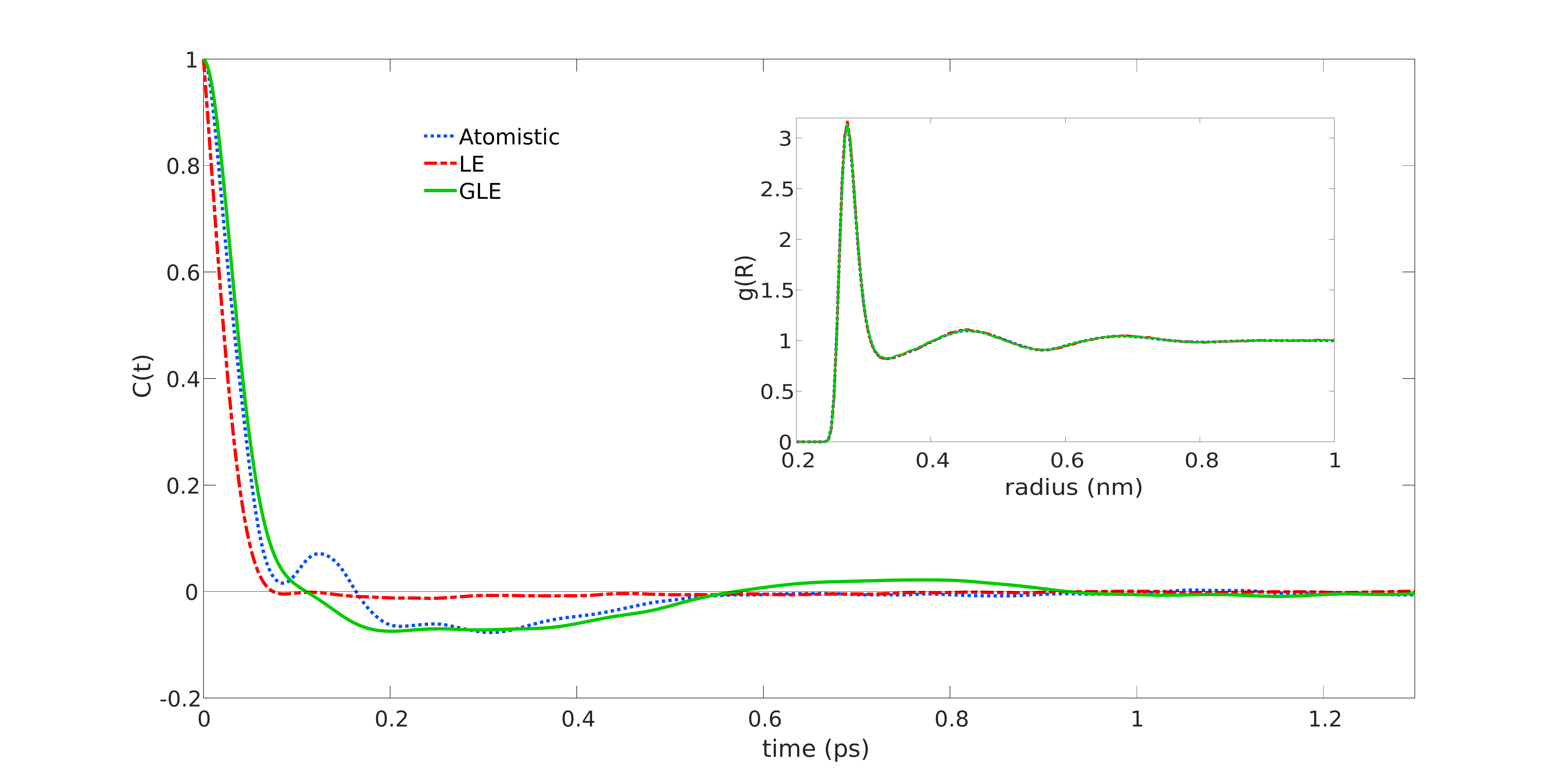}
\caption{\textit{Main panel}: Velocity autocorrelation function $C(t)$ of the water molecules' center of mass obtained in the three different setups considered in this work: ``Atomistic'' refers to the results obtained from the fully atomistic SPC-E water system simulated in the microcanonical ensemble; ``LE" refers to the results obtained from the CG model simulated via a plain Langevin equation; Finally, ``GLE" results were computed from the CG simulation that integrates the GLE with the optimised memory kernel and noise. In contrast to the CG results obtained via a plain LE, the CG system simulated via the GLE reproduces the atomistic VACF with notable accuracy. \textit{Inset}: Comparison of the radial distribution functions $g(R)$ among the water molecules' centers of mass calculated in the three different setups. We observe that the CG potential obtained through IBI reproduces the atomistic results for the $g(R)$ in both the LE and GLE cases.}
  \label{figure-Kernel_VACF}
\end{figure*}

Given the pair potential and the optimised kernel and noise, an in-house molecular dynamics integrator implementing a GLE solver was employed to perform simulations of the CG water system, see Supporting Information. We further simulated the CG model by relying on a plain Langevin equation (LE), so as to better quantify the improvement introduced by our GLE-based approach in reproducing the dynamics of the all-atom reference. Structural as well as dynamical properties of water were investigated: specifically, we computed the RDF $g(R)$ among the water molecules' centers of mass and, most importantly, their diffusion coefficient and velocity autocorrelation function (VACF). The results of these analyses in the three different setups---all-atom, CG LE and CG GLE simulations---are reported in Table~\ref{table-MSD} and Fig.~\ref{figure-Kernel_VACF}.

\begin{center}
\begin{table*}[ht]
\begin{tabular}{|p{4.5cm}|c|c|c|}
\hline
 & Atomistic & LE & GLE \\ \hline
Diffusion coefficient from MSD $[\mathrm{cm^2/s}  \times 10^{-5}]$ & $2.44 \pm 0.11$  &$2.40 \pm 0.06 $        & $2.42 \pm 0.02 $\\ \hline
Diffusion coefficient from VACF $[\mathrm{cm^2/s}  \times 10^{-5}]$ & $1.76 \pm 0.03$  & $1.99 \pm 0.04$        & $2.01 \pm 0.05$\\ \hline
Temperature {[}K{]}   & $298.09 \pm 1.18$                & $297.52 \pm 0.54 $         &$ 298.02 \pm 1.43$      \\ \hline
\end{tabular}
\caption{\textit{Top and middle row}: diffusion coefficient in the three different setups: Atomistic model, CG model simulated via a plain LE, and CG model simulated via the GLE with the optimised kernel and noise. \textit{Bottom row}: temperature of the system in the three setups. All simulations were ran at the same thermodynamical state point of volume and number of particles, see text.}
\label{table-MSD}
\end{table*}
\end{center}

First, we verified that the structural properties of the low-resolution model as obtained in both the CG setups were consistent with those of the reference, high-resolution one. That this is the case can be inferred from the inset of Fig. \ref{figure-Kernel_VACF}, where we report the RDF $g(R)$ for the three cases of interest: all-atom simulation, CG simulation through the LE, and CG simulation with the GLE. The associated curves overlap within line thickness, showing that, as expected~\cite{di2019systematic}, the structural equilibrium features of the CG system are decoupled from its kinetics.

We then moved to the computation of dynamical quantities, starting with the diffusion coefficient $D$ that was obtained from the slope of the mean square deviation (MSD) of the molecules' center of mass according to the relation MSD$(t) = 6Dt$. We observe that the coarse-grained LE and GLE results for $D$ are in good agreement with the atomistic reference, see Table~\ref{table-MSD}. Such an agreement is not surprising, as the reproduction of the all-atom diffusion coefficient was more or less explicitly enforced in the dynamics of the two CG models: more directly in the case of the LE, where we appropriately tuned the associated dissipation coefficient (see Sec.~\ref{sec:mdsim}), and somewhat implicitly in the GLE, where we introduced a constraint that relates the integral of the memory kernel to $D$ in the action minimisation workflow (see Supporting Information).

The most important dynamical property to quantify the improvement introduced by the GLE is thus the VACF $C(t)$, with
\begin{equation}
C(t)=\frac{ \sum_{t_0} \sum_{I=1}^{N_B} {\bf V}_I(t+t_0) {\bf V}_I(t_0) }{\sum_{t_0}\sum_{I=1}^{N_B} {\bf V}_I(t_0) {\bf V}_I(t_0) },
\end{equation}
where ${\bf V}_I(t)$ is the velocity of the $I^{th}$ CG water molecule center of mass at time $t$, $t_0$ being the reference starting point. Critically, this time-dependent observable was never involved in either the LE or GLE parametrisation workflow.

Fig.~\ref{figure-Kernel_VACF} shows that the VACF of the CG model simulated in the plain LE setup is markedly different from the all-atom one: more specifically, it decays to zero over significantly smaller time scales, and further fails to give rise to the negative region that is observed in the atomistic $C(t)$ before this reaches saturation. On the contrary, the CG VACF obtained by relying on the GLE with the optimised memory kernel and noise reproduces the high-resolution reference with remarkable accuracy. Interestingly, for small and intermediate time scales CG results for $C(t)$ appear to constitute a smoothened analogue of their atomistic counterpart; at the same time, we note the presence of a small, spurious oscillation that arises for long time scales in the GLE VACF and that is instead absent in the all-atom case. This comparison thus demonstrates that the proposed method allows the self-consistent parametrisation of a generalised Langevin equation that, once employed to simulate a coarse-grained model, endows it with a dynamics that is able to match the one associated to the underlying, original all-atom representation in a sensibly more quantitative manner.

It is now interesting to compute the diffusion coefficient, rather than from the mean square displacement, directly from the VACF making use of the relation \cite{Zwanzig_book}
\begin{equation}\label{vacfdiff}
D = \frac{1}{3} \int_0^\infty C(t) dt.
\end{equation}
We report the results associated to the three cases of interests in Table~\ref{table-MSD}. As for the all-atom simulation, we observe that the numerical value of the diffusion coefficient of water obtained via Eq.~\ref{vacfdiff} is different from the one derived from the mean square displacement, a discrepancy that is commonly present in the estimate of Green-Kubo-like coefficients via $C(t)$ \cite{Reilly1971VAF}. The same mismatch between MSD and VACF results for $D$ separately holds, albeit to a lesser extent, also in the case of the CG model simulated in the LE and GLE frameworks. We note that calculating the diffusion coefficients of the three different setups via the corresponding VACFs worsens the agreement between CG and all-atom predictions that was instead observed in the MSD case, see again Table~\ref{table-MSD}; at the same time, LE and GLE VACF results for $D$ are still perfectly compatible. Importantly, Fig.~\ref{figure-Kernel_VACF} displays that such an agreement between the dynamic properties of the two CG models is only apparent, as the underlying velocity autocorrelation functions substantially differ, with the LE $C(t)$ critically lacking consistency with the underlying high-resolution reference.

\section{Conclusions}\label{sec:conclusions}

In conclusion, the presented approach offers a novel conceptual framework to gain insight into the problem of the loss of dynamical consistency between high-resolution models and their coarse-grained counterpart, and thus contributes a new tool to mitigate it. In this work we introduced a time non-local action integral that depends on the coarse-grained coordinates explicitly, as well as on the residual fast coordinates in an effective, parametric manner. On the one hand, the minimisation of this action with respect to the coarse-grained variables results in the generalised Langevin equation governing the coarse-grained system; on the other hand, by computing this action on the coarse-grained trajectory obtained from a reference, all-atom simulation, one can carry out its minimisation with respect to the parameters representing the effect of the fast variables on the slow ones. This latter operation results in consistency conditions that link the time scales of the atomistic and CG models.

We followed this strategy to optimise the memory and noise components of the GLE to be employed in the simulation of a CG model. We chose a reasonable {\it Ansatz} for the kernel, and optimised its parameters through a derivative-free minimisation of the CG action. The numerical results obtained for a CG representation of a system consisting of water molecules confirmed that the theoretical framework and the proposed method substantially improve the accuracy with which the low-resolution model reproduces the velocity autocorrelation function of the all-atom reference. This improvement shows that the numerical procedure illustrated and applied in the present work can thus be fruitfully employed in the parametrisation of CG models that would reproduce not only the equilibrium, structural properties of a system, but also its dynamics.

Altogether, these results support the hypothesis that the reference system's fast dynamics, which is integrated out in the process of coarse-graining, satisfies a least action principle where the slow degrees of freedom constitute a given background. This provides comforting evidence that dynamical problems in the field of coarse-graining can be addressed in the framework of a global action, thereby contributing a novel instrument to understand the origin of the time-scale distortions that affect effective models of soft matter.

\section{Methods}\label{sec:methods}

\subsection{Coarse-grained effective potential: the iterative Boltzmann inversion method}

In principle, the multi-body potential of mean force (MB-PMF) $W({\bf R})$ acting among the CG effective interaction sites defined in Eq.~\ref{eq:consist:a} completely accounts for all equilibrium properties of the low-resolution model. In practice, however, $W({\bf R})$ can be computed explicitly only in few very simple cases \cite{diggins2018optimal}; hence, approximations are needed. Several approaches aim at approximating the MB-PMF with a sum of (state-dependent) effective pair potentials, that is,
\begin{equation}\label{eq:PMFpair}
W({\bf R}) \simeq  U({\bf R})= \sum_{I<J}U_2({\bf R}_I,{\bf R}_J),
\end{equation}
where, in homogeneous and isotropic systems,
\begin{equation}
U_2({\bf R}_I,{\bf R}_J) = U_2(|{\bf R}_I-{\bf R}_J|)=U_2(R_{IJ}). \label{eq:twobodypot}
\end{equation}
In this work, we determine the pair potential $U_2$ between two CG water beads by relying on the popular iterative Boltzmann inversion (IBI) method \cite{reith2003IBI,rosenberger2016IBI}. The IBI workflow starts from an initial guess $U_2^0$ for the pair interaction, e.g. given by
\begin{eqnarray}
U_2^{0}(R) = -k_BT\ln(g^{ref}(R)),
\end{eqnarray}
where $g^{ref}(R)$ is the reference atomistic radial distribution computed on the CG sites; the potential is then iteratively updated according to the following rule:
\begin{eqnarray}
\label{eq:IBI}
U_2^{n+1}(R) = U_2^{n}(R) + k_B T \ln \left( \frac{g^{n}(R)}{g^{ref}(R)} \right ),
\end{eqnarray}
where $g^{n}(R)$ is the radial distribution function obtained from a simulation of the CG model that employs the potential $U_2^{n}(R)$. As the number of iterations $n$ increases, $g^{n}(R)$ gets closer and closer to the reference RDF, and the update in Eq.~\ref{eq:IBI} becomes progressively smaller; the procedure is interrupted when the desired degree of convergence of the $U_2$ is achieved.
In this work, the IBI procedure was carried out using the VOTCA package \cite{ruhle2009versatile}.

\subsection{Molecular dynamics simulations}\label{sec:mdsim}
{\it Reference atomistic simulation.} MD simulations were performed for $1001$ three-points, rigid SPC/E water molecules in a cubic box with periodic boundary conditions by relying on the GROMACS 2019 software suite~\cite{van2005gromacs,abraham2015gromacs}. The Lennard-Jones intermolecular potential acting between two oxygen atoms was cut-off at $0.8$ nm, while the electrostatics of the system was computed through the Particle-mesh Ewald (PME) method. The LINCS algorithm was employed to preserve the holonomic constraints of the oxygen-hydrogen intramolecular bonds. In all simulations, the time step was set to $\Delta t=2$~fs.

A set of preliminary runs enabled the equilibration of the system around an average temperature and pressure of $T=298$~K and $p=1$~bar, respectively. Specifically, we initially performed a simulation of $0.2$~ns in the NVT ensemble by relying on the Nos\'{e}-Hoover thermostat; subsequently, a second equilibration in the NPT ensemble of $0.2$~ns, achieved by combining the Nos\'{e}-Hoover thermostat with a Parrinello-Rahman barostat, provided additional control over the system pressure. An equilibrated configuration extracted from the NPT run, whose linear box size was $L\simeq3.09$~nm, was then employed as initial condition for a microcanonical (NVE) production simulation of $0.5$~ns, where we integrated the system's equation of motion through a velocity Verlet algorithm. This simulation provided the reference data for the construction of the CG model as well as for extrapolating the thermodynamic properties of interest, namely the system temperature ($T=298$~K, see Table~\ref{table-MSD}), diffusion coefficient, velocity autocorrelation function $C(t)$, and radial distribution function $g(R)$ among the water molecules' centers of mass.

{\it CG model.} Starting from the all-atom results obtained for the $g(R)$, the effective pair potential $U_2(R)$ acting between two CG water molecules was parametrised by relying on the iterative Boltzmann inversion method (IBI) \cite{reith2003IBI,rosenberger2016IBI}, see Eqs.~\ref{eq:PMFpair}-\ref{eq:IBI}, making use of the VOTCA software \cite{ruhle2009versatile}. The initial configuration of the CG system was generated by mapping an equilibrated, all-atom NVE snapshot onto point-like beads located on the molecules' centers of mass, see Fig.~\ref{fig:water_system}. The simulations necessary for the IBI method to compute the CG potential were then conducted in NVT conditions, with $T=298$~K; in doing so, we relied on the equivalence between the canonical and microcanonical ensembles of the all-atom system, so that NVE results for the $g(R)$ are equal to their canonical counterpart when computed at the same thermodynamic state point.

{\it CG simulations.} Given the IBI pair potential $U_2(R)$ acting among water beads, simulations of the resulting CG model were performed in two different setups: a framework in which the time evolution of the low-resolution system was dictated by a plain Langevin equation, as well as one in which the GLE approach with the optimised memory kernel and noise was employed. The aim of the LE simulation was to provide a reference behaviour against which quantifying the ability of our method to mitigate the loss of dynamical consistency that arises when passing from an all-atom to a CG description of the system.

Both LE and GLE simulations involved 1001 CG beads in a cubic box of $L\simeq 3.09$~nm side with periodic boundary conditions, and were performed with a time step $\Delta t=2$~fs and at a temperature $T=298$~K. The initial configuration of the system was provided by a CG snapshot equilibrated in the NVT ensemble and extracted from the IBI procedure at convergence. The LE simulation relied on an in-house code that implemented the numerical integration scheme proposed in Ref.~\cite{goga2012efficient}, where we tuned the damping parameter so that the diffusion coefficient calculated in the CG water system coincides with its all-atom counterpart. As for the GLE, a modified velocity Verlet algorithm was employed to integrate the equations of motion, see Sec.~VI of the Supporting Information, using a in-house code available in a Zenodo repository  with DOI 10.5281/zenodo.6037951.

\subsection{Action minimisation and determination of the GLE parameters}
\label{sec:kern}

To implement the numerical minimisation of the effective CG action functional, we define the memory kernel in terms of a function depending on a set of parameters ${\bm \zeta}$ that are representative of the effect of the fast degrees of freedom as well as the ones the optimisation acts upon, see Eqs.~\ref{eq:CG_act_expl} and~\ref{eq:const-dyn}. Specifically, we define the kernel as follows:
\begin{equation}
\label{eq:kernelform}
K(t,{\bm \zeta})=K(t,a,{\bm b})=g(t,a)b(t,{\bm b})=e^{-a t^2}b(t,{\bm b}),
\end{equation}
that is, a Gaussian prefactor $g(t,a)$ followed by a modulation term $b(t,{\bm b})$, the latter being a positive function of time such that $b(t_k)=b_k $ at time $t_k=k\Delta t$. It follows that the set of parameters $\bm \zeta $ consists in the set of values $[a, b_0, b_1,..,b_{M-1}]$. The kernel in Eq. \ref{eq:kernelform} is thus implemented in the definition of the action, Eq. \ref{eq:CG_act_expl}, and the minimisation is performed to find the specific set of parameters $a$ and ${\bm b}= [b_0,b_1,...,b_{M-1}]$ that make the discretised functional stationary. In addition to the optimised kernel, knowledge of these parameters further enables the parameterisation of the noise terms in the GLE. The detailed description of this procedure, including the definition of the constraints and regularisation terms employed in its course, is provided as Supporting Information.

\section{Data availability}
This work consisted in three main computational steps: the atomistic simulation and the parametrisation of the effective potential, performed with the GROMACS software~\cite{van2005gromacs,abraham2015gromacs}; the optimisation of the memory kernel; and the LE/GLE simulations and related analyses. The last two steps were carried out making use of in-house MATLAB codes. All input files and scripts needed to obtain these data are available in the Zenodo repository with DOI 10.5281/zenodo.6037951.

\section{Acknowledgments}

The authors thank Leonardo Ricci for a critical reading and insightful comments. This project received funding from the European Research Council (ERC) under the European Union's Horizon 2020 research and innovation program (Grant 758588). RP acknowledges support from the Italian Ministry of Education, University and Research (MIUR) through the FARE grant for the project HAMMOCK (Grant R18ZHWY3NC).

\section{Author contributions}

RP conceived the study and proposed the method. PL performed the simulations, wrote the software, and collected the data. PL and RM developed the method. All authors contributed to the analysis and interpretation of the data. All authors drafted the paper, reviewed the results, and approved the final version of the manuscript.

\section{Competing interests}

The authors declare no competing interests.

\section{Supporting Information}

In the Supporting Information (\emph{i}) we report a brief summary of the Mori-Zwanzig formalism; (\emph{ii}) we prove that the effective, time non-local CG action introduced in this work, when minimised with respect to the CG trajectory, results in the GLE for the dynamics of the low-resolution system; (\emph{iii}) we provide all technical details concerning the  discretisation and minimisation of the CG action necessary to extract the optimised parameters of the GLE; (\emph{iv}) we derive the constraints on the memory kernel introduced in the minimisation workflow; and (\emph{v}) we report the numerical integrator employed in this work to solve the GLE for the CG system.

\bibliography{bibliography}

\end{document}